\documentstyle[preprint,prb,aps,fleqn]{revtex}

\begin{document}
\title{Ladder operator formalisms and generally deformed oscillator algebraic
structures of quantum states in Fock space}
\author{Xiaoguang Wang\thanks{%
email:xyw@aphy.iphy.ac.cn}}
\address{CCAST(World Laboratory),P.O.Box 8730, Beijing 100080 \\
and Laboratory of Optical Physics, Institute of Physics, \\
Chinese Academy of Sciences, Beijing 100080, People's Republic of China}
\date{\today}
\maketitle

\begin{abstract}
We show that various kinds of one-photon quantum states studied in the field
of quantum optics admit ladder operator formalisms and have the generally
deformed oscillator algebraic structure. The two-photon case is also
considered. We obtain the ladder operator formalisms of two general states
defined in the even/odd Fock space. The two-photon states may also have a
generally deformed oscillator algebraic structure. Some interesting examples
of one-photon and two-photon quantum states are given.
\end{abstract}


\section{Introduction}

The interesting quantum states of the radiation field such as coherent
states(CSs)\cite{CS}, squeezed states(SSs)\cite{SS}, binomial states(BSs)%
\cite{BS} and negative binomial states(NBSs)\cite{NBS} have been studied in
detail in the literature. The well known CS has many applications in both
quantum optics and condensed matter physics. The CS is the eigenvector of
the boson annihilation operator, while the SS is the eigenvector of a linear
combination of the boson annihilation and creation operators. The BS
interpolates between two fundamental states, the CS and the Fock state, and
reduces to them in two different limits\cite{BS}. It partakes the properties
of the CS and the Fock state. The photon number distribution of the BS is
binomial distribution in probability theory, while the photon number
distribution of the NBS is negative binomial distribution. The NBS can also
reduce to the CS in a certain limit. 

The CS has
the Heisenberg-Weyl algebraic structure while the SS has the su(1,1) Lie
algebraic structure. It is found that the BS and the NBS also have Lie
algebraic structures. The algebra involved in the BS\cite{BSfu} is the su(2)
Lie algebra via Holstein-Primakoff realization\cite{HP} and that involved in
the NBS\cite{NBSfu} is the su(1,1) Lie algebra via Holstein-Promakoff
realization. It is well known that the CS and SS can be written as the
displacement operator formalisms. The BS and NBS also admit displacement
operator formalisms. As a generalization of the BS and a generalization of
both the BS and the NBS, the hypergeometric state(HGS)\cite{HGSfu} and P${%
\acute{o}}$lya state(PS) \cite{PSfu} are introduced by Fu et al,
respectively. The HGS reduce to BS in a certain limit and the NHGS reduces
to the BS and the NBS in two different limits. The photon number
distribution of the HGS is the hypergeometric distribution in probability
theory and that of the NHGS is the P${\acute{o}}$lya distribution. The
algebraic structure of the HGS and PS are the well investigated generally
deformed oscillator (GDO) algebra\cite{GDO}. Recently, Roy et al. and Fan et
al. also introduce two kinds of quantum states which are claimed to be
BS-NBS intermediate states\cite{Roy,Fan}, but we have shown that these two
kinds of states are identical to the PSs \cite{WXG1}.

We see that every quantum state mentioned above has its own
algebraic structure. Once natural question is: what is the algebraic structure for 
a general quantum state? Usually we often study three classes of quantum
states.
One is the state with linear combination of first finite Fock states,
another is
the state without the first finite Fock states, and the last one is the state
with all Fock states. In this paper, we will investigate 
these three classes of general states and find their algebraic structures and ladder operator formalisms. 
Both one-photon and two-photon quantum states are considered.

The GDO algebra first appeared in Heisenberg's theory of nonlinear spinor dynamics%
\cite{GDO}. Many physical systems are found to posess the GDO algebra symmetry\cite
{GDO1}. The GDO albebra is an associate algebra over the complex number field $C$
with generators $A^{+},A,{\cal {N}}$ and the unit 1 satisfying
\begin{equation}
\lbrack {{\cal N},A^{+}]=A^{+},[{\cal N},A]=-A,AA^{+}=F({\cal N}+1),A^{+}A=F(%
{\cal N}),}
\end{equation}
where the Hermitian non-negative function $F$ is called the structure
function, which should satisfy the condition $F(0)=0$ in order to have Fock
representation. It will be shown that many quantum states 
bear GDO algebraic structure.

\section{One-photon quantum states}

\subsection{Quantum states as a linear combination of first finite Fock
states}

We first consider the state $|x,M\rangle $ defined as a linear combination
of first finite Fock states in an $(M+1)$-dimensional subspace of the Fock
space 
\begin{equation}
|x,M\rangle =\sum_{n=0}^MC(n,x,M)|n\rangle ,
\end{equation}
where $x$ denotes parameters of the state, $M$ is a non-negative integer and 
$|n\rangle $ is the usual Fock state. We assume the coefficients in this
paper are all non-zero. One typical example of the state $|x,M\rangle $ is
the BS. Next we want to find the ladder operator formalism of the general state $%
|x,M\rangle $.

Let the operator $f(\hat{N})a$ act on the state and we get 
\begin{equation}
f(\hat{N})a|x,M\rangle =\sum_{n=0}^{M-1}f(n)C(n+1,x,M)\sqrt{n+1}|n\rangle ,
\end{equation}
where $f(\hat{N})$ is a nonlinear function of the operator $\hat{N}=a^{+}a$.
Here $a^{+}$ and $a$ are the creation and annihilation operators of the
radiation field, respectively. If we choose 
\begin{equation}
f(\hat{N})=\frac{C(\hat{N},x,M-1)}{\sqrt{\hat{N}+1}C(\hat{N}+1,x,M)},
\end{equation}
Eq.(3) becomes 
\begin{equation}
f(\hat{N})a|x,M\rangle =|x,M-1\rangle .
\end{equation}
The operator $f(\hat{N})a$ transforms the state $|x,M\rangle $ to the state $%
|x,M-1\rangle $. We find another operator $g(\hat{N})\sqrt{M-%
\hat{N}}$ which can make this transformation.
It is easy to evaluate that 
\begin{equation}
g(\hat{N})\sqrt{M-\hat{N}}|x,M\rangle =|x,M-1\rangle ,
\end{equation}
where 
\begin{equation}
g(\hat{N})=\frac{C(\hat{N},x,M-1)}{\sqrt{M-\hat{N}}C(\hat{N},x,M)}.
\end{equation}
The operator $\sqrt{M-\hat{N}}$ removes the number state $|M\rangle $ from
the state $|x,M\rangle $ and the operator $g(\hat{N})$ makes the removed
state be the state $|x,M-1\rangle .$ Note that the operator $g(\hat{N})\sqrt{%
M-\hat{N}}$ is not equal to the operator $C(\hat{N},x,M-1)/C(\hat{N},x,M)$
since the operator $1/\sqrt{M-\hat{N}}$ is defined in the $M-$dimensional
subspace instead of ($M+1)-$dimensional subspace.

Combining Eqs.(5) and (6) leads to 
\begin{equation}
f(\hat{N})a|x,M\rangle =g(\hat{N})\sqrt{M-\hat{N}}|x,M\rangle .
\end{equation}
Substituting Eqs.(4) and (7) into the above equation , we obtain the ladder
operator formalism of the state $|x,M\rangle $ as 
\begin{equation}
\lbrack \hat{N}+\frac{(M-\hat{N})C(\hat{N},x,M)}{\sqrt{\hat{N}+1}C(\hat{N}%
+1,x,M)}a]|x,M\rangle =M|x,M\rangle .
\end{equation}

Now let us examine the algebraic structure involved in the above equation.
Define ${\cal A}$ as an associate algebra with generators 
\begin{equation}
\hat{N},A_M^{-}=\frac{(M-\hat{N})C(\hat{N},x,M)}{\sqrt{\hat{N}+1}C(\hat{N}%
+1,x,M)}a,A_M^{+}=(A_M^{-})^{\dagger }.
\end{equation}
Then it is easy to verify that these operators satisfy the commutation
relations 
\begin{equation}
\lbrack \hat{N},A_M^{\pm }]=\pm A_M^{\pm },\text{ }A_M^{+}A_M^{-}=F(\hat{N}),%
\text{ }A_M^{-}A_M^{+}=F(\hat{N}+1),
\end{equation}
where the function 
\begin{equation}
F(\hat{N})=(M-\hat{N}+1)^2\frac{C^2(\hat{N}-1,x,M)}{C^2(\hat{N},x,M)}.
\end{equation}
This algebra ${\cal A}$ is nothing but the GDO
algebra with the structure function $F(\hat{N})$. 

In terms of the generators
of the algebra ${\cal A}$, Eq.(8) can be rewritten as 
\begin{equation}
(\hat{N}+A_M^{-})|x,M\rangle =M|x,M\rangle .
\end{equation}

Below we study several interesting special cases of the state $|x,M\rangle .$

1. The BS is defined as\cite{BS} 
\begin{equation}
|\eta ,M\rangle =\sum_{n=0}^M\left[ {%
{{M} \choose {n}}%
}\eta ^n(1-\eta )^{M-n}\right] ^{1/2}|n\rangle .
\end{equation}
Here $\eta $ is a real parameter satisfying $0<\eta <1$. From Eq.(9) and
(14), we get the following ladder operator formalism for the BS

\begin{equation}
\lbrack \hat{N}+\sqrt{(1-\eta )/\eta }\sqrt{(M-\hat{N})}a]|\eta ,M\rangle
_{BS}=M|\eta ,M\rangle _{BS},
\end{equation}
which is identical to that obtained in Ref.\cite{BSfu}.

2. As a one-parameter generalization of the BS, the HGS is given by\cite
{HGSfu}

\begin{equation}
|L,M,\eta \rangle _{HGS}=\sum_{n=0}^M\left[ {%
{{L\eta } \choose {n}}%
}{%
{{L\bar{\eta}} \choose {M-n}}%
}\right] ^{1/2}{%
{{L} \choose {M}}%
}^{-1/2}|n\rangle ,
\end{equation}
where $\bar{\eta}=1-\eta $ , $L$ is a real number satisfying $L\ge \max
\{M\eta ^{-1},M\bar{\eta}^{-1}\}$, and 
\begin{equation}
{%
{{x} \choose {n}}%
}=\frac{x(x-1)...(x-n+1)}{n!},{%
{{x} \choose {0}}%
}\equiv 1.
\end{equation}

From the above equation and Eq.(9), we obtain

\begin{equation}
\lbrack \hat{N}+(\frac{L{\bar{\eta}-M+\hat{N}+1}}{L\eta -\hat{N}})^{1/2}%
\sqrt{(M-\hat{N})}a]|L,M,\eta \rangle _{HGS}=M|L,M,\eta \rangle _{HGS},
\end{equation}
which is the ladder operator formalism of the HGS.

3.As a BS-NBS intermediate state, the PS is introduced as\cite{PSfu}

\begin{equation}
|\eta ,\gamma ,M\rangle _{PS}=\sum_{n=0}^MP_n^M(\gamma ,\eta )|n\rangle ,
\end{equation}
where 
\begin{eqnarray}
P_n^M(\gamma ,\eta ) &=&{%
{{M} \choose {n}}%
}^{1/2}\{\prod_{k=1}^n[\eta +(k-1)\gamma ]\}^{1/2}\{\prod_{k=1}^{M-n}[\bar{%
\eta}+(k-1)\gamma ]\}^{1/2}  \nonumber \\
&&\{\prod_{k=1}^M[1+(k-1)\gamma ]\}^{-1/2}
\end{eqnarray}
and $\gamma >0$ is a real constant.

The ladder operator formalism of the PS is obtained as

\begin{equation}
\lbrack \hat{N}+(\frac{{\bar{\eta}+(M+\hat{N}-1)\gamma }}{\eta +\hat{N}%
\gamma })^{1/2}\sqrt{(M-\hat{N})}a]|\eta ,\gamma ,M\rangle _{PS}=M|\eta
,\gamma ,M\rangle _{PS}.
\end{equation}

4. As a reference state to detect the phase of a quantum state, Barnett and
Pegg introduced a reciprocal binomial state(RBS)\cite{RBS}

\begin{equation}
|\theta ,M\rangle _{RBS}=\left( 1/\sum_{n=0}^M{%
{{M} \choose {n}}%
}^{-1}\right) \sum_{n=0}^M{%
{{M} \choose {n}}%
}^{-1/2}\exp (in\theta )|n\rangle .
\end{equation}
The ladder operator formalism is directly obtained as

\begin{equation}
\lbrack \hat{N}+\frac{M-\hat{N}}{\hat{N}+1}\exp (-i\theta )\sqrt{(M-\hat{N})}%
a]|\theta ,M\rangle _{RBS}=M|\theta ,M\rangle _{RBS}.
\end{equation}

5.Pegg and Barnett\cite{PBPS}defined the Hermitian phase operator on a
finite-dimensional state space, which makes it possible to study the phase
properties of quantum states of a single mode of  electromagnetic field. The
Pegg-Barnett phase states (PBPS) $|\theta _m\rangle $ can be defined as 
\begin{equation}
|\theta _m,M\rangle _{PBPS}=\frac 1{(M+1)^{1/2}}\sum_{n=0}^M\exp (in\theta
_m)|n\rangle 
\end{equation}
The phase states in Eq.(24) form an orthonormal set provided that we have 
\begin{equation}
\theta _m=\theta _0+2\pi m/(s+1),m=0,1,...,s,
\end{equation}
where $\theta _0$ is an arbitrary reference phase.

We give the ladder operator formalism of the PBPS as

\begin{equation}
\lbrack \hat{N}+\frac{M-\hat{N}}{\sqrt{\hat{N}+1}}\exp (-i\theta
_m)a]|\theta _m,M\rangle _{PBPS}=M|\theta _m,M\rangle _{PBPS}.
\end{equation}

6. The generalized geometric state (GGS) is defined as\cite{GGS}

\begin{equation}
|Y,M\rangle _{GGS}=\left( \frac{1-|Y|}{1-|Y|^{M+1}}\right)
^{1/2}\sum_{n=0}^MY^{n/2}|n\rangle
\end{equation}
where $Y$ is a complex parameter.

The ladder operator formalism of the GGS is directly given as

\begin{equation}
\lbrack \hat{N}+\frac{M-\hat{N}}{\sqrt{Y}\sqrt{\hat{N}+1}}a]|Y,M\rangle
_{GGS}=M|Y,M\rangle _{GGS}.
\end{equation}

We also give the corresponding structure functions of the above six quantum
states, the BS, HGS, PS, RBS, PBPS and GGS as follows:

\begin{eqnarray}
F_{BS}(N) &=&(M-N+1)^3\frac{1-\eta }\eta ,  \nonumber \\
F_{HGS}(N) &=&(M-N+1)^3\frac{L(1-\eta )-M+N}{L\eta -N+1},  \nonumber \\
F_{PS}(N) &=&(M-N+1)^3\frac{(1-\eta )+(M+N-2)\gamma }{\eta +(N-1)\gamma }, 
\nonumber \\
F_{RBS}(N) &=&\exp (-2i\theta )(M-N+1)^5/N^2,  \nonumber \\
F_{PBPS}(N) &=&\exp (-2i\theta _m)(M-N+1)^4/N,  \nonumber \\
F_{GGS}(N) &=&(M-N+1)^4/(YN).
\end{eqnarray}

\subsection{Quantum states without first finite Fock states}

We further investigate the state $|x,M\rangle ^{-}$ defined as 
\begin{equation}
|x,M\rangle ^{-}=\sum_{n=M}^\infty D(n,x,M)|n\rangle .
\end{equation}
This state has no first finite Fock states. One type of this state is the
so-called photon-added quantum state\cite{EQS} 
\begin{equation}
|\psi ,M\rangle =N_Ma^{\dagger M}|\psi \rangle ,
\end{equation}
where $|\psi \rangle $ may be an arbitrary quantum state 
\begin{equation}
|\psi \rangle =\sum_{n=0}^\infty C(n,x)|n\rangle ,
\end{equation}
and $N_M$ is a normalization constant. For the first time, the photon-added
states were introduced by Agarwal and Tara as photon-added coherent
states(PACSs)\cite{EQS}.

Let the operators $f(\hat{N})a^{\dagger }$ and $g(\hat{N})%
\sqrt{\hat{N}-M}$ act on the state $|x,M\rangle ^{-}$ , which  leads to 
\begin{eqnarray}
f(\hat{N})a^{\dagger }|x,M\rangle ^{-} &=&|x,M+1\rangle ^{-}, \\
g(\hat{N})\sqrt{\hat{N}-M}|x,M\rangle ^{-} &=&|x,M+1\rangle ^{-},
\end{eqnarray}
where 
\begin{eqnarray}
f(\hat{N}) &=&\frac{D(\hat{N},x,M+1)}{\sqrt{\hat{N}}D(\hat{N}-1,x,M)}, \\
g(\hat{N}) &=&\frac{D(\hat{N},x,M+1)}{\sqrt{\hat{N}-M}D(\hat{N},x,M)}.
\end{eqnarray}

The operators used here, $f(\hat{N})a^{\dagger }$ and $g(\hat{N})\sqrt{\hat{N%
}-M}$ , are different from those in deriving the ladder operator formalism
of the state $|x,M\rangle $. By applying the two operators on the state $%
|x,M\rangle ^{-},$ we can transform the state $|x,M\rangle ^{-}$ to $%
|x,M+1\rangle ^{-}.$

From Eqs.(33)-(36), we obtain 
\begin{equation}
\lbrack \hat{N}-\frac{(\hat{N}-M)D(\hat{N},x,M)}{\sqrt{\hat{N}}D(\hat{N}%
-1,x,M)}a^{\dagger }]|x,M\rangle ^{-}=M|x,M\rangle ^{-}.
\end{equation}
This is the ladder operator formalism of the state $|x,M\rangle ^{-}$ in
terms of the operators $\hat{N}$ and $a^{\dagger }$. By multiplying the
annihilation operator $a$ on the above equation from left, it can be written
in terms of the operators $\hat{N}$ and $a$

\begin{equation}
(\hat{N}+1-M)a|x,M\rangle ^{-}=\frac{(\hat{N}+1-M)\sqrt{\hat{N}+1}D(\hat{N}%
+1,x,M)}{D(\hat{N},x,M)}|x,M\rangle ^{-}
\end{equation}

As a special case of the state $|x,M\rangle ^{-},$ the photon-added
state(Eq.(31)) can be expanded as 
\begin{equation}
|\psi ,M\rangle =N_M\sum_{n=M}^\infty C(n-M,x)[n!/(n-M)!]^{1/2}|n\rangle .
\end{equation}
From the above equation and Eq.(37), we obtain the ladder operator formalism
of the photon-added state as 
\begin{equation}
\lbrack \hat{N}-\frac{C(\hat{N}-M,x)}{C(\hat{N}-M-1,x)}\sqrt{\hat{N}-M}%
a^{\dagger }]|\psi ,M\rangle =M|\psi ,M\rangle .
\end{equation}

The above equation can be written in terms of the operators $\hat{N}$ and $%
a, $

\begin{equation}
(\hat{N}+1-M)a|\psi ,M\rangle =\frac{C(\hat{N}+1-M,x)}{C(\hat{N}-M,x)}\sqrt{%
\hat{N}+1-M}(\hat{N}+1)]|\psi ,M\rangle .
\end{equation}

The algebra involved in the state $|x,M\rangle ^{-}$ is also the GDO algebra 
${\cal B}$, with the generators  
\begin{equation}
\hat{N},B_M^{+}=\frac{(\hat{N}-M)D(\hat{N},x,M)}{\sqrt{\hat{N}}D(\hat{N}%
-1,x,M)}a^{\dagger },B_M^{-}=(B_M^{+})^{\dagger }.
\end{equation}
These satisfy the commutation relations 
\begin{equation}
\lbrack \hat{N},B_M^{\pm }]=\pm B_M^{\pm },B_M^{+}B_M^{-}=G(\hat{N}%
),B_M^{-}B_M^{+}=G(\hat{N}+1)
\end{equation}
with the structure function 
\begin{equation}
G(\hat{N})=(\hat{N}-M)^2\frac{D^2(\hat{N},x,M)}{D^2(\hat{N}-1,x,M)}.
\end{equation}
In terms of the generators of the algebra ${\cal B}$, Eq.(37) becomes 
\begin{equation}
\lbrack \hat{N}-B_M^{+}]|x,M\rangle ^{-}=M|x,M\rangle ^{-}.
\end{equation}

As an example, we derive the ladder operator formalism for the PACS. The CS
expanded in Fock space is

\begin{equation}
|\psi \rangle _{CS}=\exp (-|\alpha |^2/2)\sum_{n=0}^\infty \frac{\alpha ^n}{%
\sqrt{n!}}|n\rangle ,
\end{equation}
where $\alpha $ is a complex number. Using the coefficients of the CS and
Eq.(41), we find the ladder operator formalism for the PACS to be

\begin{equation}
(\hat{N}+1-M)a|\alpha ,M\rangle =\alpha (\hat{N}+1)|\alpha ,M\rangle .
\end{equation}
Since the operator 1/($\hat{N}+1$) is non-zero in the whole Fock space, we
get

\begin{equation}
\lbrack 1-M/(\hat{N}+1)]a|\alpha ,M\rangle =\alpha |\alpha ,M\rangle .
\end{equation}
According to the definition of the nonlinear coherent states (NLCSs)\cite
{NLCS}

\begin{equation}
f(\hat{N})a|\alpha \rangle _{NLCS}=\alpha |\alpha \rangle _{NLCS,}
\end{equation}
so the PACS is an NLCS as discussed by Sivakumar\cite{Sivakumar}. Here $f(%
\hat{N})$ is a nonlinear function of $\hat{N}.$

As another example, we consider a new definition of NBS(NNBS)\cite{NNBS}
recently introduced by Barnett: 
\begin{equation}
|\eta ,M\rangle _{NNBS}=\sum_{n=M}^\infty \left[ {%
{{n} \choose {M}}%
}\eta ^{M+1}(1-\eta )^{n-M}\right] ^{1/2}|n\rangle ,
\end{equation}
It is found that the NNBS and the BS have similar properties if the roles of
the creation operator $a^{\dagger }$ and annihilation operator $a$ are
interchanged.

Form Eq.(38), the ladder operator formalism of the NNBS is obtained as

\begin{equation}
\lbrack \sqrt{\hat{N}+1-M}/(\hat{N}+1)]a|\eta ,M\rangle _{NNBS}=\sqrt{1-\eta 
}|\eta ,M\rangle _{NNBS}.
\end{equation}
As seen from the above equation, the NNBS is an NLCS with the nonlinear
function $f(\hat{N})=\sqrt{\hat{N}+1-M}/(\hat{N}+1)$. Eq.(51) was obtained
in one of our previous papers\cite{WXG2}

\subsection{General quantum states}

By an analogous method to that used in obtaining Eq.(37), we obtain the
ladder operator formalism of the general state $|\psi \rangle ($Eq.(32)$)$
as 
\begin{equation}
\lbrack \hat{N}-\frac{C(\hat{N},x)}{C(\hat{N}-1,x)}\sqrt{\hat{N}}a^{\dagger
}]|\psi \rangle =0.
\end{equation}
Multiplying both sides of the above equation by $a$ from the left, we get
another form of the ladder operator formalism of the state $|\psi \rangle $, 
\begin{equation}
a|\psi \rangle =\frac{C(\hat{N}+1,x)}{C(\hat{N},x)}\sqrt{\hat{N}+1}|\psi
\rangle .
\end{equation}
The algebra involved in the state $|\psi \rangle $ is also the GDO algebra
with generators 
\begin{equation}
\hat{N},[C(\hat{N},x)/C(\hat{N}-1,x)]\sqrt{\hat{N}}a^{\dagger },\text{ }a%
\sqrt{\hat{N}}C(\hat{N},x)/C(\hat{N}-1,x).
\end{equation}
The corresponding structure function is $\hat{N}^2C^2(\hat{N},x)/C^2(\hat{N}%
-1,x)$.

Using the coefficients of the CS (Eq.(46)), we obtain the well-known ladder
operator formalism of the CS from Eq.(53):

\begin{equation}
a|\alpha \rangle _{CS}=\alpha |\alpha \rangle _{CS}.
\end{equation}
This equation can be naturally obtained by letting $M=0$ in Eq.(48).

Now we consider the geometric state(GS) which is defined as\cite{GS}

\begin{equation}
|\eta \rangle _{GS}=\eta ^{1/2}\sum_{n=0}^\infty (1-\eta )^{n/2}|n\rangle 
\end{equation}
This is also called the Susskind-Glogower phase state, phase eigenstate,and
coherent phase state\cite{GS} in the literature. From Eq.(53), we get the
ladder operator formalism of the GS

\begin{equation}
a|\eta \rangle _{GS}=\sqrt{1-\eta }\sqrt{N+1}|\eta \rangle _{GS}.
\end{equation}

Since the operator $\sqrt{N+1}$ is not zero in the whole Fock space, we
obtain

\begin{equation}
a/\sqrt{N+1}|\eta \rangle _{GS}=\sqrt{1-\eta }|\eta \rangle _{GS}.
\end{equation}

This equation shows that the GS is an NLCS with the nonlinear function $1/%
\sqrt{N+1}.$ By setting $M=0$ in Eq.(51), Eq.(51) reduces to Eq.(58). This
fact is easily understood since the NNBS can be viewed as photon-added GS%
\cite{WXG2}.

Further we study another example, the NBS, which is defined as\cite{NBS}

\begin{equation}
|\eta ,M\rangle _{NBS}=\sum_{n=0}^\infty (1-\eta )^{M/2}{%
{M+n-1 \choose n}%
}^{1/2}\eta ^{n/2}|n\rangle .
\end{equation}

From Eq.(53), we get

\begin{equation}
\frac 1{\sqrt{M+\hat{N}}}a|\eta ,M\rangle _{NBS}=\eta ^{1/2}|\eta ,M\rangle
_{NBS},
\end{equation}
which shows that the NBS is an NLCS with the nonlinear function $f(\hat{N}%
)=1/\sqrt{M+\hat{N}}$. This result is identical to that obtained by us before%
\cite{WXG3}.

It is interesting to investigate the Kerr state (KS) which is defined as\cite
{KS}

\begin{equation}
|\alpha ,\theta \rangle _{KS}=\exp (-|\alpha |^2/2)\sum_{n=0}^\infty \frac{%
\alpha ^n\exp [-i\theta n(n-1)]}{\sqrt{n!}}|n\rangle .
\end{equation}
From Eq.(53), the ladder operator formalism of the KS is obtained as

\begin{equation}
\exp (-2i\hat{N}\theta )a|\alpha ,\theta \rangle _{KS}=\alpha |\alpha
,\theta \rangle _{KS}.
\end{equation}
When $\theta =0,$ Eq.(62) naturally reduce to Eq.(55). Eq.(62) show that the
KS is an NLCS with the nonlinear function $f(\hat{N})=\exp (-2i\hat{N}\theta
).$ 

Here we give a more general form of Eq.(9) as
\begin{equation}
\left( \sqrt{\eta }\hat{N}+\sqrt{1-\eta }f(\hat{N})a\right) |\eta ,\alpha \rangle
=\alpha |\eta ,\alpha \rangle ,  \label{Eigenequ}
\end{equation}
where the parameter $0<\eta <1$ is introduced as discussed in Ref.\onlinecite{Fu2000}.
The parameter $\alpha$ is the eigenvalue to be determined.
According to the procedure same as that by Fu et al. \cite{Fu2000}, we can obtain
a new type of quantum state, i.e., the intermediate number-nonlinear coherent states.
If we choose $f(\hat{N})=\exp (-2i\hat{N}\theta)$, we obtain
the interesting intermediate number-Kerr state. The detailed discussion will be given elsewhere.

All the states discussed in this section are of the one-photon type.
Two-photon quantum states will be studied in the next section.

\section{Two-photon quantum states}

The representative two-photon states are the squeezed vacuum states(SVSs)
(defined below) and even/odd coherent states(ECS/OCS)\cite{EOCS}. These
states are defined in either even Fock space or odd Fock space.

The squeezed vacuum state is defined as

\begin{equation}
|\xi \rangle _{SVS}=S(\xi )|0\rangle ,
\end{equation}
where $S(\xi )=\exp (\xi K_{+}-\xi ^{*}K_{-})$ is the squeezing operator and 
$\xi =r\exp (i\theta )$. Here $K_{+}=a^{+2}/2$ and $K_{-}=a^2/2.$ The two
operators together with the operator $K_0=\hat{N}/2+1/4$ form a su(1,1) Lie
algebra.

The representation of su(1,1) on the usual Fock space is completely
reducible and decomposes into a direct sum of the even Fock space ($S_0$)
and odd Fock space ($S_1$)\cite{PRAfu},

\begin{equation}
S_j=\text{span}\{||n\rangle _j\equiv |2n+j\rangle |n=0,1,2,...\},\text{ }%
j=0,1.
\end{equation}

Representations on $S_j$ can be written as

\begin{eqnarray}
K_{+}||n\rangle _j &=&\sqrt{(n+1)(n+j+1/2)}||n+1\rangle _j, \\
K_{-}||n\rangle _j &=&\sqrt{(n)(n+j-1/2)}||n-1\rangle _j,  \nonumber \\
K_0||n\rangle _j &=&(n+j/2+1/4)||n\rangle _j.  \nonumber
\end{eqnarray}
The Bargmann index $k=1/4$($3/4$) for even(odd) Fock space.

It is easily seen that the SQS is defined in the even Fock space. To obtain
the expansion of the SQS in the even Fock space, we use the decomposed form
of the squeezing operator

\begin{equation}
S(\xi )=\exp (e^{i\theta }\tanh r\text{ }a^{+2}/2)(\cosh
r)^{-(a^{+}a+1/2)}\exp (-e^{-i\theta }\tanh r\text{ }a^2/2).
\end{equation}

From the above equation and Eq.(63), we obtain the expansion as

\begin{equation}
|\xi \rangle _{SVS}=(\cosh r)^{-1/2}\sum_{n=0}^\infty \sqrt{(2n)!}%
[e^{i\theta }\tanh r/2]^n/n!||n\rangle _0.
\end{equation}

Now we consider two general states in even/odd Fock space

\begin{eqnarray}
|x\rangle _e &=&\sum_{n=0}^\infty C_e(n,x)||n\rangle _0, \\
|x\rangle _o &=&\sum_{n=0}^\infty C_o(n,x)||n\rangle _{1.}
\end{eqnarray}

For convenience, we introduce the number operators $\hat{N}_{0,}\hat{N}_1$
defined by

\begin{eqnarray}
\hat{N}_0 &=&K_0-1/4,\hat{N}_0||n\rangle _0=n||n\rangle _0, \\
\hat{N}_1 &=&K_0-3/4,\hat{N}_1||n\rangle _1=n||n\rangle _1.
\end{eqnarray}

Using the same method as that used in deriving Eq.(37), we obtain the ladder
operator formalisms of the states $|x\rangle _{e/o},$

\begin{eqnarray}
\lbrack \hat{N}_0-\frac{\sqrt{\hat{N}_0}C_e(\hat{N}_0,x)}{C_e(\hat{N}_0-1,x)%
\sqrt{\hat{N}_0-1/2}}\frac{a^{+2}}2]x\rangle _e &=&0, \\
\lbrack \hat{N}_1-\frac{\sqrt{\hat{N}_1}D_o(\hat{N}_1,x)}{D_o(\hat{N}_1-1,x)%
\sqrt{\hat{N}_1+1/2}}\frac{a^{+2}}2]x\rangle _o &=&0.
\end{eqnarray}

The above equation can be written in terms of $a^2$ and $N_j$,

\begin{eqnarray}
\lbrack \frac{C_e(\hat{N}_0+1,x)\sqrt{\hat{N}_0+1/2}(N_0+1)}{C_e(N_0,x)}-%
\sqrt{\hat{N}_0+1}\frac{a^2}2]x\rangle _e &=&0, \\
\lbrack \frac{C_o(\hat{N}_1+1,x)\sqrt{N_1+3/2}(N_1+1)}{C_o(N_1,x)}-\sqrt{%
\hat{N}_1+1}\frac{a^2}2]x\rangle _o &=&0.
\end{eqnarray}

Substituting the coefficients of the SVS into Eq.(72) leads to

\begin{equation}
\frac 1{\hat{N}+1}a^2|\xi \rangle _{SVS}=\exp (i\theta )\tanh r|\xi \rangle
_{SVS}.
\end{equation}

We see that the SVS is the two-photon nonlinear coherent state(TPNLCS) which
can be defined as\cite{TPNLCS}

\begin{equation}
f(\hat{N})a^2|\alpha \rangle _{TPNLCS}=\alpha |\alpha \rangle _{TPNLCS}.
\end{equation}

As another squeezed state, the squeezed first excited state(SFES) is defined
as

\begin{equation}
|\xi \rangle _{SFES}=S(\xi )|1\rangle ,
\end{equation}
By using Eq.(66), we get the expansion of the state $|\xi \rangle _{SFES}$

\begin{equation}
|\xi \rangle _{SFES}=(\cosh r)^{-3/2}\sum_{n=0}^\infty \sqrt{(2n+1)!}%
[e^{i\theta }\tanh r/2]^n/n!||n\rangle _0.
\end{equation}
Similar to the derivation of Eq.(76), we obtain

\begin{equation}
\frac 1{\hat{N}+2}a^2|\xi \rangle _{SVS}=\exp (i\theta )\tanh r|\xi \rangle
_{SVS}
\end{equation}
$.$\newline
The above equation shows that the SFES is a TPNLCS.

Other examples are the even CS(ECS) and odd CS(OCS)\cite{EOCS}

\begin{eqnarray}
|\alpha \rangle _{ECS} &=&1/\sqrt{\cosh |\alpha |^2}\sum_{n=0}^\infty \alpha
^{2n}/\sqrt{(2n)!}||n\rangle _{0,}. \\
|\alpha \rangle _{OCS} &=&1/\sqrt{\sinh |\alpha |^2}\sum_{n=0}^\infty \alpha
^{2n+1}/\sqrt{(2n+1)!}||n\rangle _{1.}
\end{eqnarray}

The ladder operator formalism can be easily obtained as for Eqs(74) and (75)

\begin{eqnarray}
a^2|\alpha \rangle _{ECS} &=&\alpha ^2|\alpha \rangle _{ECS}, \\
a^2|\alpha \rangle _{OCS} &=&\alpha ^2|\alpha \rangle _{OCS}.  \nonumber
\end{eqnarray}
This is just what we expected.

As seen from Eqs.(72) and (73), we conclude that the algebra involved in the
two general two-photon states is the GDO algebra with generators

\begin{equation}
\hat{N}_0,\frac{\sqrt{\hat{N}_0}C_e(\hat{N}_0,x)}{C_e(\hat{N}_0-1,x)\sqrt{%
\hat{N}_0-1/2}}\frac{a^{+2}}2,\frac{a^2}2\frac{\sqrt{\hat{N}_0}C_e(\hat{N}%
_0,x)}{C_e(\hat{N}_0-1,x)\sqrt{\hat{N}_0-1/2}},
\end{equation}
and

\begin{equation}
\hat{N}_1,\frac{\sqrt{\hat{N}_1}D_o(\hat{N}_1,x)}{D_o(\hat{N}_1-1,x)\sqrt{%
\hat{N}_1+1/2}}\frac{a^{+2}}2,\frac{a^2}2\frac{\sqrt{\hat{N}_1}D_o(\hat{N}%
_1,x)}{D_o(\hat{N}_1-1,x)\sqrt{\hat{N}_1+1/2}}.
\end{equation}
The corresponding structure functions are

\begin{eqnarray}
F_e(\hat{N}_0) &=&\frac{\hat{N}_0^2C_e^2(\hat{N}_0,x)}{C_e^2(\hat{N}_0-1,x)},
\nonumber \\
F_o(\hat{N}_1) &=&\frac{\hat{N}_1^2D_o^2(\hat{N}_1,x)}{D_o^2(\hat{N}_1-1,x)}.
\end{eqnarray}

\section{Conclusions}

We have shown that various kinds of states in the field of quantum optics
admit ladder operator formalisms and have the GDO algebraic structures. The
corresponding structure functions have been obtained. As examples we have  
given ladder operator formalisms for the BS, RBS, NBS, NNBS, HGS, PS, RBS, 
PBPS, GS, GGS and KS. We have also considered the two-photon case and get  
the ladder operator formalisms for two general states defined in the
even/odd Fock space . We find that the algebra involved in the two general
states is also GDO algebra. The ladder operator formalisms of some special
two-photon states, the SVS, SFES, ECS and OCS, have been obtained. The SVSs
and SFESs are found to be TPNLCSs. In addition, the method proposed here can 
be applied to the study of ladder operator formalism and the algebraic
structure of quantum states for su(1,1) and su(2) Lie algebra.

Some quantum states discussed in the present paper seems only for academic interestes. However with the
development of the quantum state engineering technique\cite{StateEngi}, these states can be
fabricated in principle.

\vspace{2cm}

{\bf Acknowledgments}: The author is grateful for discussions with Professor
H. C. Fu and the help of Professors C. P. Sun, S. H. Pan and G. Z. Yang. This work
was partially supported by the National Science Foundation of China with
Grant No. 19875008.

\end{document}